\documentclass[twocolumn,showpacs,preprintnumbers,amsmath,amssymb]{revtex4-1}

\usepackage{graphicx}
\usepackage{dcolumn}
\usepackage{bm}

\begin{document}

\title{Dynamical approach to heavy-ion induced fission using actinide target nuclei 
at energies around the Coulomb barrier$^{*}$}

\author{Y.~Aritomo$^{1,2}$ , K.~Hagino$^{3}$, K.~Nishio$^{1}$, and S.~Chiba$^{1}$}

\affiliation{$^{1}$ Japan Atomic Energy Agency, Tokai, Ibaraki, 319-1195, Japan}%
\affiliation{$^{2}$ Flerov Laboratory of Nuclear Reactions, JINR, Dubna, 141980, Russia}%
\affiliation{$^{3}$ Department of Physcis, Tohoku University, Sendai 980-8578, Japan}%




\begin{abstract}
In order to describe heavy-ion fusion reactions around the Coulomb barrier with an actinide target nucleus,
we propose a model which combines the coupled-channels approach and a fluctuation-dissipation
model for dynamical calculations.
This model takes into account
couplings to the collective states of the interacting nuclei
in the penetration of the Coulomb barrier and the subsequent dynamical
evolution of a nuclear shape from the contact configuration.
In the fluctuation-dissipation model with a Langevin equation,
the effect of nuclear orientation at the initial impact on the prolately deformed target nucleus
is considered.
Fusion-fission, quasi-fission and deep quasi-fission are separated
as different Langevin trajectories on the potential energy surface.
Using this model, we analyze the experimental data for the mass distribution of
fission fragments (MDFF) in the reactions of $^{34,36}$S+$^{238}$U and $^{30}$Si+$^{238}$U
at several incident energies around the Coulomb barrier.
We find that the time scale in the quasi-fission as well as the deformation of fission fragments
at the scission point
are different between the $^{30}$Si+$^{238}$U and $^{36}$S+$^{238}$U systems, causing
different mass asymmetries of the quasi-fission.

\vspace{0.5cm}
$^{*}$ {\footnotesize This is the simplified version excluded figures with large file size, which could not be uploaded here.
To obtain the complete version (included 21 figures), please contact to aritomo24@muj.biglobe.ne.jp. }

\end{abstract}




\maketitle



\section{Introduction}

The prediction of the existence of the ``Island of Stability" in the nuclear chart has
encouraged searches of new elements \cite{myer66}.
The synthesis of these superheavy elements has been carried out using heavy-ion fusion
reactions between stable nuclei, in which two different types of reaction have been employed.
In the cold fusion reactions, lead and bismuth targets are used \cite{ogan75,hofm00}.
The element with $Z=113$ reported by RIKEN used this type of reaction \cite{mori03}.
The superheavy nuclei (SHN) synthesized in the cold fusion reaction
produce nuclei with relatively small number of neutrons.
The other type of reaction, called the hot fusion reaction, on the other hand,
uses actinide nuclei as targets.
With this type of reaction, production of elements with $Z = 114, 115, 116, 117$ and 118
were reported by the Flerov Laboratory of Nuclear Reactions (FLNR) \cite{ogan01}.
These nuclei as well as those produced as descendants in the $\alpha$-decay chain have
relatively larger number of neutrons.
Recently, other laboratories than FLNR also performed experiments of hot fusion reactions
and obtained results that are consistent with the data by FLNR \cite{hofm07,stav09,elli10,dull10}.
At present, attempts to produce elements 119 and 120 are made or planed
in several facilities using actinide targets.
In order to produce new nuclei or elements not discovered so far,
an accurate prediction of the production cross sections is an important
issues in the SHN research.

An important quantity for a prediction of the cross section is the probability of fusion
after the interacting nuclei have the initial contact.
Due to the complexity of the process, however, a good method to predict the fusion probability
has not been established well.
Furthermore, actinide nuclei are prolately deformed, so that one needs to introduce nuclear
orientation as an additional degree of freedom.
The effect of nuclear orientation on fusion probability has been evident already
in the reactions using rare-earth nuclei with a prolate shape \cite{mits00,nish00}.
This makes the hot fusion in contrast to the cold fusion reactions, which use spherical nuclei.

Our strategy to calculate the fusion probability is to use the unified model \cite{zagr05},
which has been developed by the FLNR theory group \cite{zagr05,zagr07a,zagrt}.
The model can describe every entrance and exit channels in heavy-ion collisions,
and can calculate also the time evolution of the nuclear shape,
where the binary decay of the composite system (fission) can be treated.
In this model, a system first feels a diabatic potential in the early stage of the initial collision.
The potential is then gradually shifted to an adiabatic potential in a time-dependent manner.
A trajectory calculation is performed on the time-dependent unified potential energy
surface by using the Langevin equation.
In the trajectory analysis, different types of fission can be separated, that is,
fusion-fission (FF) and quasi-fission process (QF).
The fusion-fission
is a fission of a compound nucleus, and is defined as the case when the trajectory enters
the region of compound-nucleus, followed by fission.
The quasi-fission, on the other hand, is the fission event whose trajectory does not enter the region
of compound nucleus.
The fusion probability is defined as the FF events normalized to all the fission events (that is, FF+QF).
The model should be constrained or checked by experiments by investigating whether
the calculation can
reproduce the measured fission spectra such as mass and total kinetic energy distributions.

Recently, the mass distributions of fission fragments (MDFF) for the reactions
$^{36,34}$S+$^{238}$U and $^{30}$Si+$^{238}$U at several incident energies around the Coulomb
barrier were
measured by the Japan Atomic Energy Agency (JAEA) \cite{nish08,nish08a,nish10a,nish10b}.
One of the findings in the experiment is that the mass asymmetry in QF
is different between the $^{30}$Si+$^{238}$U and $^{36,34}$S+$^{238}$U systems at low incident energies.
In this work, we attempt to analyze these data in order to understand the reaction mechanism.

In the previous paper \cite{arit09}, to estimate the capture and fusion cross sections,
we only considered the spherical-spherical configuration as the first approximation,
limiting to the energy region above the Bass barrier.
In order to extend the calculation down to energies below the barrier, the effect of nuclear
orientation has to be taken into account.
Such effect has been well established in the approaching phase of the reaction
using the coupled-channels approach \cite{NBT86,RHT01,HR04}.
However,
it is still difficult to calculate the adiabatic potential energy
to be used in the unified model
with the two-center parametrization
for subsequent shapes of the nuclear system, starting from the configuration of arbitrarily oriented two
deformed nuclei touching each other to the spherical compound nuclei.
In this paper, we propose a new model which can avoid this difficulty.
In the new model, all the orientation angles are
effectively taken into account by introducing the effective charge-center
distance at the contact point as a variable.
With this prescription, a dynamical calculation for superheavy elements is possible for the first
time at energies below the Coulomb barrier.

The paper is organized as follows.
In Sec.~II, we detail the framework of the new model, which combines the coupled-channels method
and the dynamical Langevin calculation.
In Sec.~III, we show the results for the cross sections and MDFF at several incident energies
for the reactions of $^{36,34}$S + $^{236}$U and $^{30}$Si + $^{236}$U.
In Sec.~IV, we discuss the reasoning of the different shapes of MDFF observed in these  reactions.
In Sec.~V, we present a summary of this study and further discussion.

\section{Model}


\subsection{Coupled-channels method}

Excitations of the rotational states in a deformed nucleus in the approaching phase of heavy-ion collisions
considerably modify the fusion barrier
and increase the capture cross sections at the sub-barrier energies.
For a heavy deformed nucleus considered in this paper,
it is reasonable to introduce the sudden approximation to the coupled-channels (CC) equations.
In this approximation,
the capture cross section is given as \cite{NBT86,RHT01,HR04},
\begin{equation}
\sigma_{\rm cap}(E)=\int^1_0d(\cos\theta)\sigma_{\rm cap}(E_{\rm cm};\theta),
\end{equation}
where $\theta$ is the angle of the incident projectile nucleus with respect to
the symmetry axis of the deformed target, and
$E_{\rm cm}$ denotes the incident energy in the center-of-mass frame.
Here, we have assumed that the target nucleus has an axially symmetric shape.
$\sigma_{\rm cap}(E_{\rm cm};\theta)$ is the capture cross section
for a given value of $\theta$,
calculated with an angle dependent internucleus potential $V_{\rm CC}(r,\theta)$,
%
\begin{widetext}
\begin{eqnarray}
V_{\rm CC}(r,\theta)&=&V^{(N)}(r,\theta)+V^{(C)}(r,\theta), \\
V^{(N)}(r,\theta)&=&\frac{-V_0}{1+\exp[(r-
R-R_T\beta_2Y_{20}(\theta)-R_T\beta_4Y_{40}(\theta))/a_{\rm WS}]},
\label{VN}
\\
V^{(C)}(r,\theta)&=&\frac{Z_PZ_Te^2}{r}
+\sum_{\lambda=2,4}\left(\beta_\lambda
+\frac{2}{7}\sqrt{\frac{5}{\pi}}\beta_2^2\delta_{\lambda,2}\right)
\,\frac{3Z_PZ_Te^2}{2\lambda+1}\frac{R_T^\lambda}{r^{\lambda+1}}
Y_{\lambda0}(\theta).
\end{eqnarray}
\end{widetext}
Here, $R_T$ is the equivalent sharp surface radius
of the target nucleus,
and $\beta_2$ and $\beta_4$ stand for the quadrupole and hexadecapole deformation
parameters of the target nucleus, respectively.
We have assumed the Woods-Saxon function for the nuclear potential, $V^{(N)}$.
Using the penetration probability for the $\ell$-th partial wave, $T_\ell$,
the capture cross sections are given by
%
\begin{equation}
\sigma_{\rm cap}(E_{\rm cm};\theta)=\frac{\pi}{k^2}\sum_{\ell=0}^\infty(2\ell+1)
T_\ell(E_{\rm cm};\theta),
\end{equation}
where $k$ is the wave number of the incident flux.
The vibrational excitations in the projectile nucleus are also taken
into account for each $\theta$ in the same way as in the computer
code {\tt CCFULL} \cite{ccfull}.

The fusion cross section is calculated by multiplying the
probability to form a compound nucleus, $P_{\rm CN}$, to the capture
probability,
$T_\ell(E_{cm};\theta)$,
at each incident angle $\theta$ and integrating it over the solid angle as
\begin{equation}
\sigma_{\rm fus}(E_{\rm cm})=\int^1_0d(\cos\theta)\sigma_{\rm fus}(E_{\rm cm};\theta),
\end{equation}
with
\begin{equation}
\sigma_{\rm fus}(E_{\rm cm};\theta)=\frac{\pi}{k^2}\sum_{\ell=0}^\infty(2\ell+1)
T_\ell(E_{\rm cm};\theta)P_{\rm CN}(E_{\rm cm},\ell,\theta).
\end{equation}
As we explain in the next subsection, the formation probability
$P_{\rm CN}$ is estimated with the dynamical calculation using
the Langevin equation.
In order to introduce the orientation angle dependence
to $P_{\rm CN}(E,\ell,\theta)$,
we make the following approximation.
We first notice that
the charge-center distance at the
touching point
is strongly correlated with
the incident angle $\theta$.
In our model, the touching distance is assumed to be
\begin{equation}
z_{\rm touch}(\theta)=
R+R_T\beta_2Y_{20}(\theta)+R_T\beta_4Y_{40}(\theta),
\label{ztouch}
\end{equation}
where $R$ is the radius parameter in the Woods-Saxon potential, Eq. (\ref{VN}).
In the dynamical calculation, we use an angle independent potential energy surface,
but starting from the angle dependent initial touching distance given by Eq. (\ref{ztouch}).
With this prescription, we can
for the first time extend the dynamical Langevin calculation down to the
subbarrier region.

\subsection{Dynamical calculation}

After the projectile enters with
an arbitrary orientation relative to the symmetry
axis of the deformed target nucleus, the collision is replaced by the one from the polar-side
of the target nucleus and the trajectory calculation starts from the configuration corresponding
to the touching distance $z_{\rm touch}$.
That is, we consider only the
nose-to-nose configuration.
In this stage, we assume that the potential has been shifted
to the adiabatic potential from the diabatic one.
In the reactions of $^{238}$U,
the static deformation of $\beta_{2}=0.215$ ($\delta \sim 0.2 $) \cite{moll95} is
used.

The nuclear shape is defined by the two-center
parametrization \cite{maru72,sato78}, which has three deformation parameters, $z_0, \delta$, and $\alpha$.
$z_{0}$ is the distance between two potential centers,
while
$\alpha=(A_{1}-A_{2})/(A_{1}+A_{2})$
is the mass asymmetry of the
colliding nuclei, where
$A_{1}$ and $A_{2}$ denote the mass numbers of heavy and light nuclei, respectively \cite{arit04}.
$\delta$ denotes the deformation of the fragments, and
is defined by
$\delta=3(R_\parallel-R_\perp)/(2R_\parallel+R_\perp)$, where $R_\parallel$ and $R_\perp$
are the half length of the axes
of an ellipse in the $z_{0}$ and $\rho$ directions of the cylindrical coordinate,
respectively, as shown in Fig.~1 in Ref. \cite{maru72}.
We assume that each fragment has the same deformation.
The deformation parameters $\delta$ and $\beta_{2}$ are related to each other as
\begin{equation}
\beta_{2}=\frac{\delta}{\sqrt{\frac{5}{16\pi}}(3-\delta)}.
\end{equation}
Notice $\delta < 1.5$, because  $R_\parallel > 0$ and $R_\perp > 0 $.
In order to reduce the computational time, we employ the coordinate $z$ defined as
$z=z_{0}/(R_{CN}B)$, where $R_{CN}$ denotes the radius of a spherical compound nucleus
and $B$ is defined as $B=(3+\delta)/(3-2\delta)$.

The neck parameter $\epsilon$ entering in the two-center parametrization has been adjusted
in Ref. \cite{zagr07} to reproduce the available data,
assuming different values between the entrance and the exit channels of the reactions.
In our study, we use $\epsilon=1$ for the entrance channel and $\epsilon=0.35$ for the exit channel.
We assume the following time dependence for the adiabatic potential \cite{karp07},
expressed in terms of the relaxation time $\tau_{\epsilon}$ for $\epsilon$:
\begin{equation}
V(q,t)=V(q,\epsilon=1)f_{\epsilon}(t)+V(q,\epsilon=0.35)[1-f_{\epsilon}(t)],
\end{equation}
with
\begin{equation}
f_{\epsilon}(t)= \exp\left(-\frac{t}{\tau_{\epsilon}}\right).
\label{epsi}
\end{equation}
Here, $q=\{z,\delta,\alpha\}$ is the deformation coordinate.
We use $\tau_{\epsilon}=10^{-20}$ sec., as it gives a reasonable account for the available data
of mass, angular, and kinetic energy distributions of outgoing fragments produced in
heavy-ion collisions (that is, deep-inelastic reaction, nuclear transfer reactions, fusion-fission)
\cite{arit09,zagr05,zagr07a,zagrt}.

For a given value of $\epsilon$ and a temperature of a system, $T$,
the adiabatic potential energy is defined as
\begin{equation}
V(q,\ell,T)=V_{\rm LD}(q)+\frac{\hbar^{2}\ell(\ell+1)}{2I(q)}+V_{\rm SH}(q,T),
 \label{vt1}
\end{equation}
\begin{equation}
V_{\rm LD}(q)=E_{\rm S}(q)+E_{\rm C}(q),
\end{equation}

\begin{equation}
V_{\rm SH}(q,T)=E_{\rm shell}^{0}(q)\Phi (T),
\end{equation}

\begin{equation}
\Phi (T)=\exp \left(-\frac{aT^{2}}{E_{\rm d}} \right).
\end{equation}
Here, $V_{\rm LD}$ is the potential energy calculated with the finite-range liquid drop model,
given as a sum of of the surface energy $E_{\rm S}$ \cite{krap79} and the Coulomb energy $E_{\rm C}$.
$V_{\rm SH}$ is the shell correction energy evaluated for each temperature using the factor $\Phi (T)$,
in which $E_{\rm d}$ is the shell damping energy chosen to be 20 MeV \cite{igna75} and
$a$ is the level density parameter.
At the zero temperature ($T=0$), the shell correction energy reduces to $E_{\rm shell}^{0}$.
The second term on the right hand side
of Eq. (\ref{vt1}) is the rotational energy
for an angular momentum $\ell$ \cite{arit04},
with a moment of inertia, $I(q)$.



Since we employ the CC model for the approaching phase
to describe the penetration of the Coulomb barrier, the two-body part of the unified mode
is omitted \cite{zagr05,zagr07a,zagrt,arit09} in our calculations.
The multidimensional Langevin equations \cite{arit04,zagr05} are thus simplified as
\begin{eqnarray}
&&\frac{dq_{i}}{dt}=\left(m^{-1}\right)_{ij}p_{j},\nonumber \\
&&\frac{dp_{i}}{dt}=-\frac{\partial V}{\partial q_{i}}
                 -\frac{1}{2}\frac{\partial}{\partial q_{i}}
                   \left(m^{-1}\right)_{jk}p_{j}p_{k} \nonumber \\
&&~~~~~~~~~~~~~~~~~~~~~-\gamma_{ij}\left(m^{-1}\right)_{jk}p_{k}\nonumber
                  +g_{ij}R_{j}(t),
\end{eqnarray}
where
$p_{i} = dq_{i}/dt$ with $i = \{z, \delta, \alpha\}$.
The summation is performed over repeated indices.
In the Lengevin equation,
$m_{ij}$ and $\gamma_{ij}$ are the shape-dependent collective inertia parameter and the
friction tensor, respectively.
The wall-and-window one-body dissipation
\cite{bloc78,nix84,feld87}is adopted for the friction tensor.
A hydrodynamical inertia tensor is adopted with the Werner-Wheeler approximation
for the velocity field \cite{davi76}.
The normalized random force $R_{i}(t)$ is assumed to be white noise, {\it i.e.},
$\langle R_{i}(t) \rangle$=0 and $\langle R_{i}(t_{1})R_{j}(t_{2})
\rangle = 2 \delta_{ij}\delta(t_{1}-t_{2})$.
The strength of the random force $g_{ij}$ is given by $\gamma_{ij}T=\sum_{k}
g_{ij}g_{jk}$.

The temperature $T$ is calculated from the intrinsic energy
of the composite system as $E_{\rm int}=aT^{2}$, where
$E_{\rm int}$ is calculated at each step of a trajectory calculation as
\begin{equation}
E_{\rm int}=E^{*}-\frac{1}{2}\left(m^{-1}\right)_{ij}p_{i}p_{j}-V(q,\ell,T=0).
\end{equation}
The excitation energy of the compound nucleus $E^{*}$ is given by $E^{*}=E_{\rm cm}-Q$, where $Q$ denotes
the $Q$-value of the reaction.

The fusion probability $P_{\rm CN}$ in Eq.~(7) is determined in our model calculation by
identifying the different trajectories on the deformation space.
It is equivalent to the number of trajectories of compound-nucleus fission
normalized to all the fission events.
Formation of the compound nucleus is defined as the case that
a trajectory
enters in a compact-shape region in the adiabatic potential energy surface.
We define the compound nucleus region (that is, the fusion box) by referring to
the ridge of the fission barrier in the coordinate space \cite{arit04}.

\section{Mass distribution of fission fragments and cross sections} 

\subsection{Reaction of $^{36,34}$S+$^{238}$U}

\begin{figure}
\centerline{
\includegraphics[height=.44\textheight]{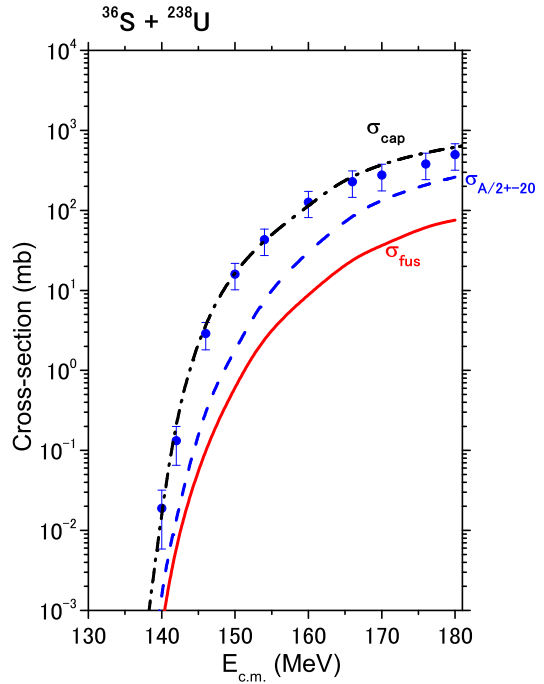}}
  \caption{(Color online) Excitation functions of $\sigma_{\rm cap}$, $\sigma_{A/2\pm
 20}$ and $\sigma_{\rm fus}$ for the $^{36}$S+$^{238}$U reaction. The experimental data
of $\sigma_{\rm fiss}$, denoted by the circles, are taken from Ref. ~\cite{nish08}.}
\label{fig_cr36s}
\end{figure}

Recently MDFF and the fission cross sections ($\sigma_\textrm{fiss}$) for the the
$^{36,34}$S + $^{238}$U reactions were measured by the JAEA group \cite{nish08,nish10a}.
In the experiment, fission events were selected in which
the momentum of projectile is fully transferred to the composite system.
For these systems, the fission cross sections are almost equal to
those of the projectiles captured inside
the Coulomb barrier, $\sigma_\textrm{cap}$.

We first analyze these systems with the new model proposed in the previous section.
The dashed-dot curve in Fig. ~\ref{fig_cr36s} shows the
calculated capture cross sections based on the coupled-channels model
for the the reaction $^{36}$S + $^{238}$U
as a function of the incident
energy.
The experimental data are taken from Ref. \cite{nish08}.
As in Refs. \cite{nish06,nish04,hind95,nish08}, we use
the computer code {\tt CCDEGEN} \cite{hagi99} by taking into account
the static deformation for $^{238}$U with
($\beta_{2}$, $\beta_{4}$)= (0.275,0.05).
Couplings to the $2^{+}$ vibrational state at 3.29 MeV in $^{36}$S (with $\beta_{2}=$0.61 \cite{rama87})
and the $3^{-}$ state at 0.73 MeV in $^{238}$U (with $\beta_{3}=0.086$ \cite{spea89}) are also considered.
We use $V_0$=105.0 MeV, $R$=10.92 fm, $R_T$=7.44 fm, and $a_{\rm WS}=$0.75 fm for the Woods-Saxon
potential, $V^{(N)}(r,\theta)$, given by Eq. (\ref{VN}).
One can see that this calculation
reproduces the measured cross sections
down to the lowest incident energy below the Bass barrier ($V_\textrm{Bass}= 158.8$ MeV) \cite{bass74}.

The solid curve in Fig.~\ref{fig_cr36s} shows the fusion cross section
$\sigma_\textrm{fus}$ obtained by the new model with CC and the Langevin equation.
At the Bass barrier, we obtain $P_{\rm CN}=0.03$ for $\ell=0$ and $\theta=0$.
The dashed line shown in Fig.~\ref{fig_cr36s} denotes the cross section
$\sigma_{A/2 \pm20}$, which is derived from the yield of the fission fragments
whose mass number is located within $\pm$20 around the symmetric fission $A_\textrm{CN}/2$.
Notice that the fusion cross sections $\sigma_\textrm{fus}$ are significantly
smaller than $\sigma_{A/2 \pm20}$. This indicates that
the mass symmetric fission does not necessarily originate from the compound-nucleus state.


In the previous study \cite{arit09},
we started the dynamical calculation with
the spherical-spherical configuration
in order to estimate the fission cross sections of the various kinds.
Because the calculation using the Langevin equation is a classical one, this did not allow us
to calculate the cross sections below the barrier.
With the new approach,
by considering the nuclear shapes at the contact configuration
for each orientation, we can now obtain the cross sections also below the Bass barrier region.

%
%

Above the Bass barrier region, the calculated $\sigma_\textrm{fus}$ with the present model
is about 7 times larger than that in the previous study \cite{arit09}.
In the present model, the Langevin calculation is started at the touching point
assuming all the kinetic energy has $-z$ direction,
while in the previous study, we started the Langevin calculation from a sufficiently large distance
between the target and projectile and the orientation dependence was not considered.
The previous results of smaller fusion probabilities are due to the loss of the kinetic energy
in the approaching process by the friction.



\begin{figure}
\centerline{
\includegraphics[height=.75\textheight]{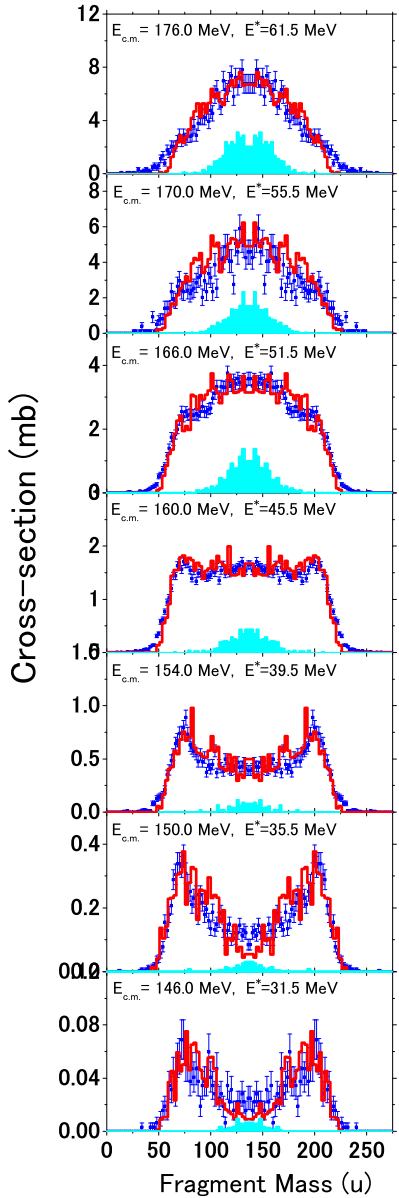}}
  \caption{(Color online) Mass distributions of fission fragments
for the reaction of $^{36}$S+$^{238}$U.
The experimental data and the calculated results are denoted
by the circles~\cite{nish08} and the histograms, respectively.
The shaded areas show the calculated fusion-fission events.}
\label{fig_mass36s}
\end{figure}

The results for the MDFF for the reaction of $^{36}$S + $^{238}$U are compared
with the experimental data \cite{nish08} in Fig.~\ref{fig_mass36s} at seven incident energies from $E_\textrm{cm}$=148.0 ($E^{*}$=31.5) MeV to $E_\textrm{cm}$=176.0 ($E^{*}$=61.5) MeV (see the histograms).
At high incident energies,
the mass distribution
has a Gaussian-like shape centered at the symmetric mass division,
whereas the mass-asymmetric fission fragments dominate at low incident energies.
The mass-asymmetric fission produces nuclei in the vicinity of the
doubly-closed shell nuclei, $^{208}$Pb and $^{78}$Ni.
The trend of the experimental data, {\it i.e.}, the incident energy dependence of MDFF,
is well reproduced by the calculation. The mass-asymmetry with $A_\textrm{H}=$ 200
at sub-barrier energies is also well reproduced.

In Fig.~\ref{fig_mass36s} we also plot the fusion-fission events by the filled histograms.
Apparently, the compound-nucleus fission has a mass-symmetric shape,
and the observed mass-asymmetric fission dominated at the low incident energies is
classified as QF.
The strong
energy dependence
of the MDFF
can be understood in terms of the orientation effect
on the fusion and QF.
The collision on the polar side have a
large probability to disintegrate as QF,
whereas the collision on the equatorial side have a larger fusion probability.
The standard deviation of the spectrum for FF is $\sigma_{m}$=37.2 u
at $E_\textrm{c.m.}$=176.0 MeV.
This value is far smaller than that of the experimental spectrum,
which also shows the Gaussian-like shape, indicating that there is a significant
contribution of QF even at this low energy.
The calculation also suggests that the measured mass-symmetric fission fragment
has another origin than the compound nucleus fission.
Such an event is defined as a deep quasi-fission process (DQF) as discussed in Ref. \cite{arit04}.

\begin{figure}
\centerline{
\includegraphics[height=.44\textheight]{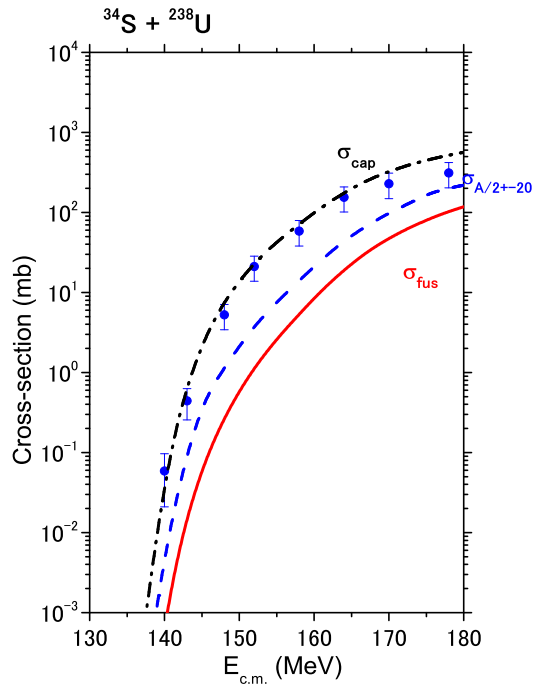}}
  \caption{(Color online)
Same as Fig. 3, but for the $^{34}$S+$^{238}$U reaction.
The experimental data are taken from Ref. ~\cite{nish10a}.}
\label{fig_cr34s}
\end{figure}

\begin{figure}
\centerline{
\includegraphics[height=.75\textheight]{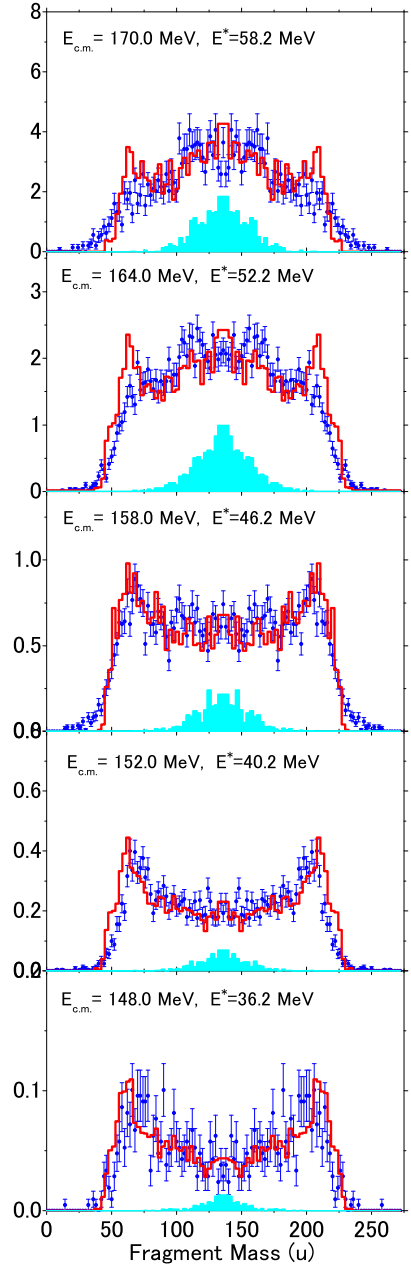}}
  \caption{(Color online)
Same as Fig. 4, but for the $^{34}$S+$^{238}$U reaction.
The experimental data are taken from Ref. ~\cite{nish10a}.}
\label{fig_mass34s}
\end{figure}

The results for the
$^{34}$S + $^{238}$U reaction are shown in Figs.~\ref{fig_cr34s} and \ref{fig_mass34s}.
In addition to the effects of the static deformation of $^{238}$U and the couplings to the 3$^{-}$ state
in $^{238}$U, we also take into account the $2^{+}$ state at 2.13 MeV in $^{34}$S \cite{fire98}
in the CC calculation.
Figure~\ref{fig_mass34s} shows the MDFF for five incident energies between
$E_\textrm{c.m}$=148.0 ($E^{*}$=36.2) MeV and $E_\textrm{c.m}$=170.0 ($E^{*}$=58.2) MeV.
The Bass barrier for this reaction is $V_\textrm{Bass}= 161.1$ MeV.
The meaning of each curve in Figs.~\ref{fig_cr34s} and \ref{fig_mass34s} is the same as in
Figs.~\ref{fig_cr36s} and \ref{fig_mass36s}.
The energy dependence of MDFF is qualitatively the same as
in the $^{36}$S + $^{238}$U reaction, and the calculations again reproduce well the experimental data.

\subsection{Reaction of $^{30}$Si+$^{238}$U}

\begin{figure}
\centerline{
\includegraphics[height=.44\textheight]{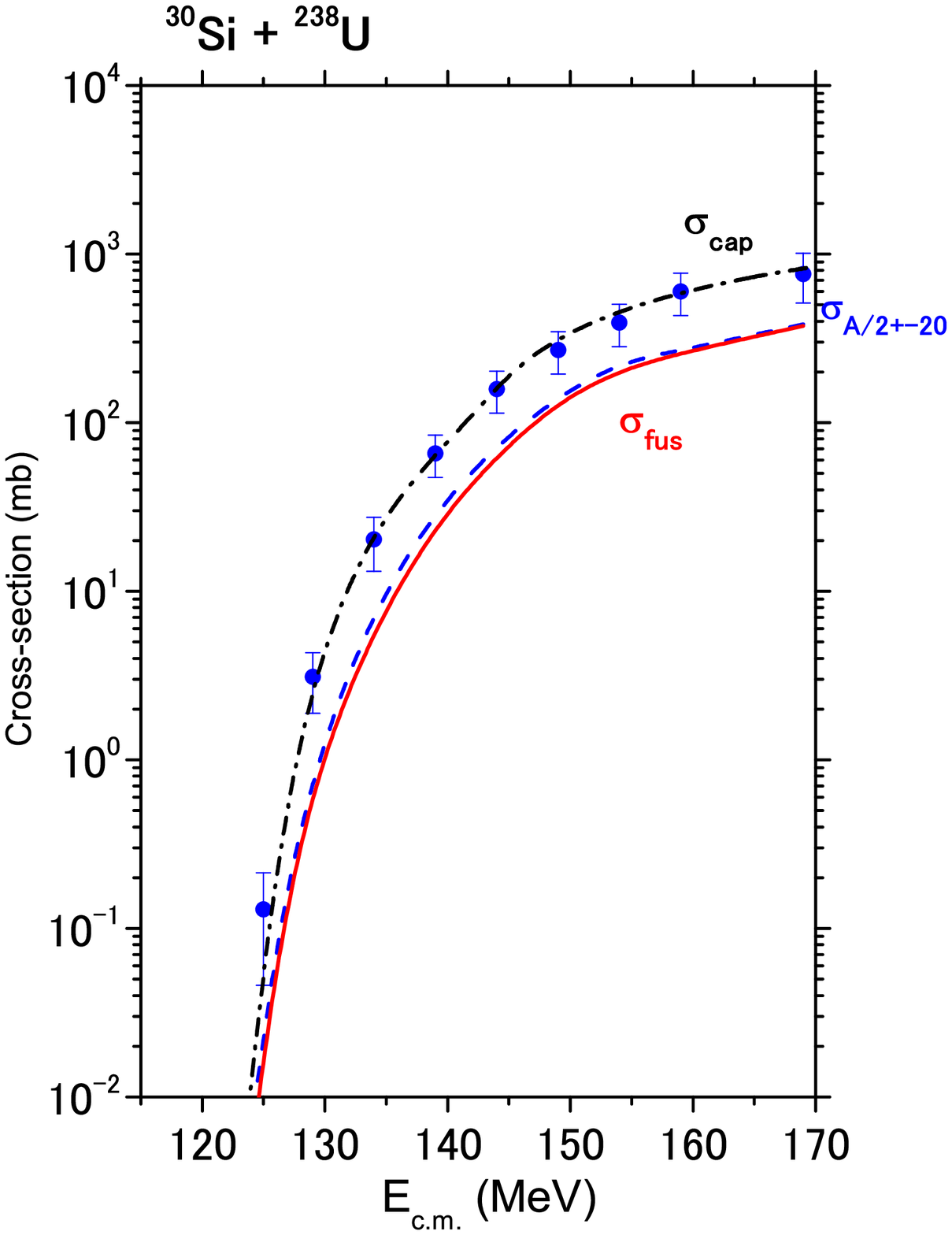}}
  \caption{(Color online)
Same as Fig. 3, but for the $^{30}$Si+$^{238}$U reaction.
The experimental data are taken from Ref. ~\cite{nish10b}. }
\label{fig_cr30si}
\end{figure}

We next analyze the $^{30}$Si+$^{238}$U reaction, for which
the MDFF and the fission cross sections have been measured by the JAEA group \cite{nish10b}.
The present calculation for the capture cross section $\sigma_\textrm{cap}$,
fusion cross section $\sigma_\textrm{fus}$ and the mass-symmetric fission cross section
$\sigma_{A/2 \pm20}$ are shown
in Fig.~\ref{fig_cr30si}
by the dashed-dot, solid and dashed curves, respectively.
For the capture cross sections, similar parameters for CC
are used as in the $^{34,36}$S + $^{238}$U reactions, except that
the couplings to the vibrational state in $^{30}$Si are not taken into account.
The present calculation for the capture cross sections
reproduces quite well the fission cross sections
down to the sub-barrier region below the Bass barrier, $V_\textrm{Bass}=$ 141.1 MeV,
better than the previous model calculation \cite{arit09}.
The predicted fusion cross sections $\sigma_\textrm{fus}$ are
about 5 to 7 times larger than the previous model \cite{arit09}
due to the same reason as in the
$^{34,36}$S+$^{238}$U reactions discussed in the previous subsection.

\begin{figure}
\centerline{
\includegraphics[height=.75\textheight]{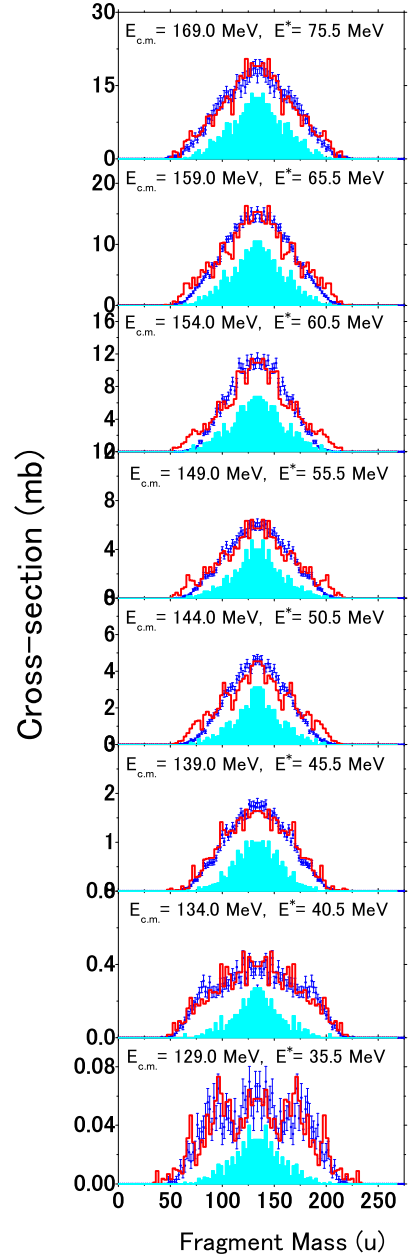}}
  \caption{(Color online)
Same as Fig. 4, but for the $^{30}$Si+$^{238}$U reaction.
The experimental data are taken from Ref. ~\cite{nish10b}. }
\label{fig_mass30si}
\end{figure}

The calculated results for MDFF
are shown in Fig.~\ref{fig_mass30si} by the histograms
for eight incident energies $E_\textrm{c.m}$ ($E^{*}$) between 129.0 (35.5) MeV and 169.0 (75.5) MeV.
We use the same parameters for the trajectory calculation
as those for the $^{36,34}$S + $^{238}$U reactions.
A significant difference in the measured
MDFF between the $^{30}$Si + $^{238}$U and the $^{34,36}$S + $^{238}$U reactions
is the mass-asymmetry in QF at subbarrier energies.
The mass-asymmetry for the $^{30}$Si + $^{238}$U system
is $A_\textrm{H}/A_\textrm{L} \sim 178/90$ as shown in
Fig. \ref{fig_mass30si}, whereas $A_\textrm{H}/A_\textrm{L}$ is $204/68$ and $200/74$
for the $^{34}$S + $^{238}$U and the $^{36}$S + $^{238}$U reactions, respectively (see Figs.
\ref{fig_mass36s} and \ref{fig_mass34s}).
The QF fragments $A_\textrm{H}/A_\textrm{L}$ = 178/90 do not fit any shell closure of neutron rich nuclei.
A production of these nuclei is therefore not associated with the local minimum of the potential
energy attained by a larger binding energy of the nascent fission fragments at the scission point,
but should originate from dynamical aspects in heavy-ion induced fission.
This feature is well reproduced by the present calculation,
in the entire energy range from above-barrier to sub-barrier.
The standard deviation of the measured MDFF decreases
from the highest incident energy down to the low energy of $E_\textrm{c.m.}=$ 139.4 MeV,
then the value suddenly increases at the sub-barrier energies of 134.0 and 129.0 MeV.
These trends are also reproduced by the calculation.

The calculated FF events are shown by the filled histograms in Fig.~\ref{fig_cr30si}.
It has a mass-symmetric distribution, and the standard deviation does not change
through all the energy points.
Therefore, the sudden broadening of the measured MDFF
at sub-barrier energies
can be attributed to the enhancement of
the mass-asymmetric QF at $A_\textrm{H}/A_\textrm{L}$ = 178/90.

\section{Analysis of the structure in the mass distribution of fission fragments}


In the previous section, we have shown that the new model
nicely reproduces the energy as well as the system dependence
of MDFF.
In this section, we discuss the origin for the difference
in the mass-asymmetry in QF between the
$^{30}$Si + $^{238}$U and $^{36}$S + $^{238}$U systems
based on our model.

\subsection{Potential energy surface along the scission line}

\begin{figure}
\centerline{
\includegraphics[height=.44\textheight]{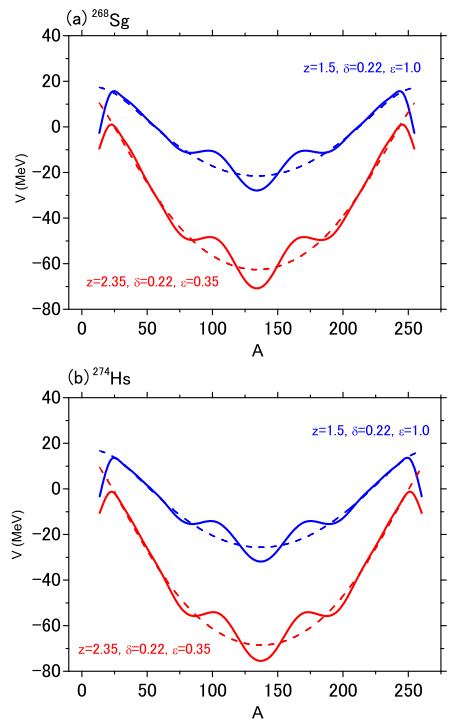}}
  \caption{(Color online) Adiabatic potential energy surfaces
near the scission point
for the nuclei $^{268}$Sg (a) and $^{274}$Hs (b).
$V_{\rm LD}$ and $V_{\rm LD}+E_{\rm shell}^{0}$ with $\delta=0.22$
are represented by the dashed and solid lines, respectively.
The red and blue lines denote the potential with
$\epsilon=0.35 (z=2.35, \delta=0.22)$ and $\epsilon=1.0 (z=1.5, \delta=0.22)$,
respectively. }
\label{fig_1dim-pot}
\end{figure}

\begin{figure}
\centerline{
\includegraphics[height=.3\textheight]{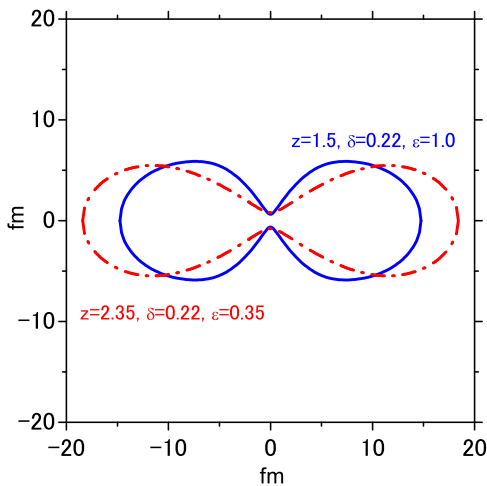}}
  \caption{(Color online) The nuclear shapes near the scission point.
  The dashed-dot and solid lines represent the nuclear shapes
corresponding to the two curves in Fig. 9, that is,
$\epsilon=0.35$ $(z=2.35, \delta=0.22)$ and
$\epsilon=1.0$ $(z=1.5, \delta=0.22)$, respectively. }
\label{fig_shape1}
\end{figure}

Let us first consider the potential energy surface for the two systems.
Generally, the shape of MDFF from a compound nucleus is
affected by the landscape of the potential energy surface,
especially near the fission saddle point and
the scission point \cite{wilk76,moll74}.
On the other hand, for QF,
the saddle point of the composite system is not passed
by the system during the evolution of nuclear shape.
A plausible explanation for the mass-asymmetry in QF
may have something to do with
the potential energy at the scission point
as the exit point of the reaction process.
Therefore, we focus on the landscape of the potential
energy surface near the scission line.

Figures~\ref{fig_1dim-pot}(a) and (b) show the adiabatic
potential energy surface near the scission point
for the nuclei $^{268}$Sg ($^{30}$Si + $^{238}$U)
and $^{274}$Hs ($^{36}$S + $^{238}$U), respectively.
The dashed and solid curves correspond to
the potential $V_{\rm LD}$ and $V_{\rm LD}+E_{\rm shell}^{0}$
with deformation of $\delta=0.22$, respectively.
This value of the deformation parameter $\delta$ is chosen
since it reproduces well the available fission data
as discussed in Refs. \cite{arit052,arit06}.
As we mentioned in Sec. II B,
the value of $\epsilon$ evolves from $\epsilon=1$ towards $\epsilon=0.35$
as a system approaches the compound-nucleus shape and fissions.
The blue and the red lines show the potential surfaces for the
configurations corresponding to
these values of $\epsilon$ ($\epsilon$=1 and 0.35), that is,
($z$, $\delta$)=(1.5, 0.22) and ($z$, $\delta$)=(2.35, 0.22), respectively.
The solid and dashed-dot curves in Fig.~\ref{fig_shape1} represent the
corresponding nuclear shapes with a mass symmetry $\alpha=$0.
Notice that for a faster process like QF,
the system does not have enough time to reach $\epsilon=0.35$
and $\epsilon$ is maintained about 1.0 still at the scission point.

Even though the shell energy $E_{\rm shell}^{0}$
has a strong local minimum
at $A \sim 208$ (which corresponds to $^{208}$Pb)
owing to the strong shell structure,
the local minimum of $V_{\rm LD}+E_{\rm shell}^{0}$ appears at $A \sim 186$
due to the parabolic shape of $V_{\rm LD}$ around $A\sim 135$ for
both $^{268}$Sg and $^{274}$Hs.
One also finds another minimum at
$A \sim 85$ for both the nuclei.
If the mass division were determined at the scission point,
the mass-asymmetric QF should thus have appeared at $A\sim 86$ and 188
for the reaction $^{36}$S + $^{238}$U and at $A\sim 84$ and 184
for the reaction $^{30}$Si + $^{238}$U.
These are not realized in the measured MDFF, however.
The two peaks observed in the experiment are located
at approximately $A \simeq 74$ and 200 in the reaction
of $^{36}$S + $^{238}$U.
Evidently, a prediction based only on the potential energy surface
at the scission line is insufficient to explain the mass
asymmetry in QF, indicating that the dynamical process plays an essential role
to describe QF.


\subsection{Analysis of the reaction dynamics using probability distribution}

\begin{figure}
\centerline{
\includegraphics[height=.3\textheight]{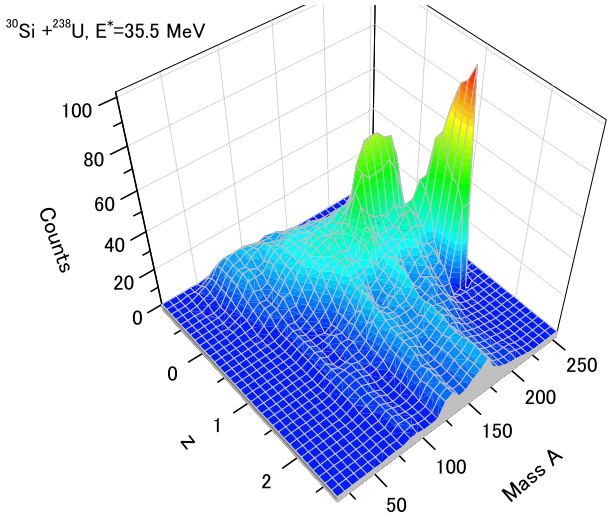}}
  \caption{(Color online) The probability distribution
constructed as an ensemble of Langevin trajectories
on the $z-A$ plane
for the reaction $^{30}$Si + $^{238}$U at $E^{*}=35.5$ MeV with $\ell=0,$ $\theta=0$. }
\label{tden106aa}
\end{figure}

We next discuss the
probability distribution of the system in the deformation space.
To this end,
we segment the coordinate space with
$\Delta z=0.10, \Delta \delta=0.06$ and $\Delta \alpha=0.06$.
We define the distribution as an ensemble of Langevin trajectories.
That is, we follow a trajectory as a function of time, and we increase
the event number at each segment when the trajectory passes through that segment.
By generating many trajectories,
we construct a distribution of events on the deformation space.


Figure~\ref{tden106aa} shows the distribution so constructed
on the $z-A$ plane for the reaction $^{30}$Si + $^{238}$U
at $E^{*}=35.5$ MeV with $\ell=0,$ and  $\theta=0$.
We consider the time evolution
of the nuclear shapes as well as the mass number only for one of the fragments, that is, the one
corresponding to the target-like nucleus.
Even though this distribution is not a probability distribution, because it is not normalized
in the whole space,
we call the distribution ``probability distribution".
From such probability distribution shown in Fig.~\ref{tden106aa},
we can understand the overall trend of the dynamical evolution of nuclear shape,
where the vertical axis of the figure means the number of events entering in the $z-A$ plane.

Two remarkable peaks are visible in Fig.~\ref{tden106aa}.
The peak located around $(z,A) \sim (1.3,238)$ corresponds to the touching point.
All the trajectories start at this point, so that the number of events at this segment is large.
The broadly distributed events in the smaller-$z$ region
correspond to the trajectories trapped in
the large-$\delta$ region as seen in Figs.\ref{fig_2dim106aa} (a) and (c).
They are the trajectories that enter into a compact region defined by $z < 0.5$ and
$75 < A < 150$, which correspond to $\sigma_{\rm fus}$ in Fig.~\ref{fig_cr30si}.
We also see the yields which leads to scission at $A \sim 178$ and 135 at $z = 2.5$,
which correspond to the peaks of MDFF shown in Fig.~\ref{fig_mass30si}.
The peak at $A \sim 178$ originates from QF, while
the peak at $A \sim 135$ originates largely from the FF, but also involves the component from DQF.


\begin{figure}
\centerline{
\includegraphics[height=.3\textheight]{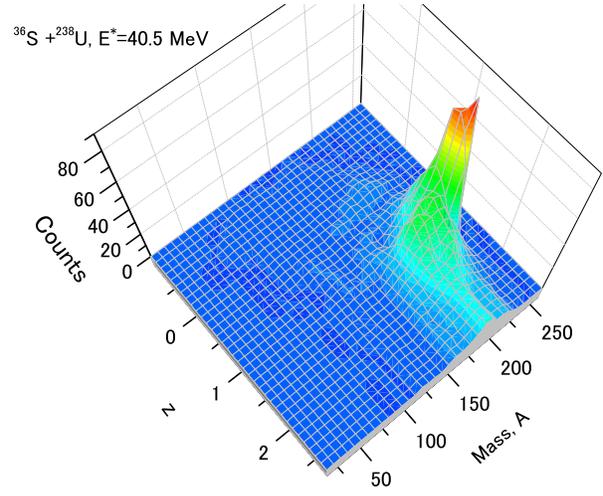}}
  \caption{(Color online)
Same as Fig. 14, but for
for the $^{36}$S + $^{238}$U reaction at $E^{*}=40.5$ MeV with $\ell=0,$ $\theta=0$.}
\label{tden108aa}
\end{figure}

The probability distribution for the $^{36}$S + $^{238}$U reaction at $E^{*}=39.5$ MeV
is shown in Fig.\ref{tden108aa}. The spectrum is dominated by the mass-asymmetric QF.
Similarly to the $^{30}$Si+$^{238}$U reaction, the large peak corresponds to the
touching point.
Almost all the trajectories move quickly to the direction of binary decay as QF,
and those which approach the compact nuclear shape with small $z$ value, corresponding to $\sigma_{\rm fus}$
in Fig.~\ref{fig_cr36s}, are significantly diminished.
QF generates fragments around $A \sim 200$ (at $z = 2.5$) in this calculation, explaining the
observed mass asymmetry of MDFF in Fig~\ref{fig_mass36s} at the lowest incident energy.

\begin{figure}
\centerline{
\includegraphics[height=.4\textheight]{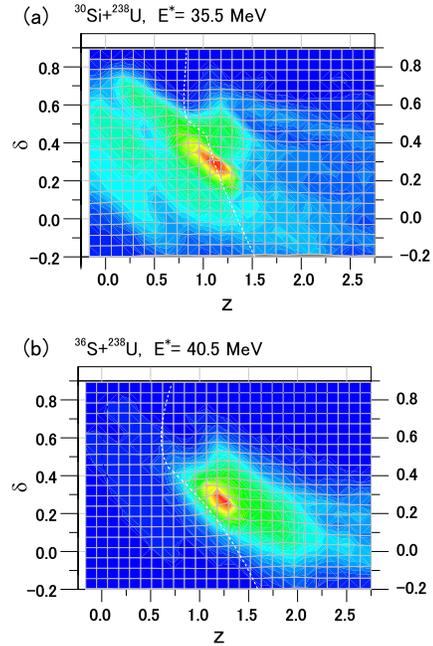}}
  \caption{(Color online) The contour maps for the probability distributions on the the $z-\delta$ plan,
  for the reactions of
  $^{30}$Si + $^{238}$U at $E^{*}=35.5$ MeV (a) and $^{36}$S + $^{238}$U at $E^{*}=40.5$ MeV (b)
with $\ell=0,$ $\theta=0$. The dashed curves correspond to the ridge lines.
 }
\label{2tden106d108d}
\end{figure}

The probability distributions projected onto the $z-\delta$ plane are
shown in Fig.~\ref{2tden106d108d}(a) and (b) for the reactions of
$^{30}$Si + $^{238}$U and $^{36}$S + $^{238}$U, respectively.
The dashed curves denote the ridge lines.
The positions at the strongest yield correspond to the touching points.
In Fig.~\ref{2tden106d108d}(a), the distribution crosses the ridge line and moves to the
large-$\delta$ region.
It spreads in the small-$z$ region and $\delta$ value distribute widely from $-$0.1 to 0.7.
After surmounting the ridge line and entering in the compact nuclear shape region, a
mono-nucleus with small $z$-shape is formed, but its deformation fluctuates significantly
due to the thermal fluctuation.
The distribution for $z < 0.5$ and $\delta < 0.2$ corresponds to the formation of
the compound nucleus.
The $^{36}$S + $^{238}$U reaction shown in Fig.~\ref{2tden106d108d}(b) has a much smaller chance to form
the mono-nucleus as compared to the reaction $^{30}$Si + $^{238}$U,
because of the smaller probability to overcome the the ridge line.
Instead, the system quickly disintegrates as QF by forming fission fragments.

We can clearly observe that the fission fragments for the reaction $^{30}$Si + $^{238}$U are
deformed with $-0.2 < \delta < 0.5$, because they include various events originating from QF, DQF and FF.
On the other hand, for the $^{36}$S + $^{238}$U reaction, the value of the deformation parameter of
the fission fragments are restricted mainly to $-0.1 < \delta < 0.2$,
primarily because QF is the dominant channel.

From the analysis of the probability distribution, it is clear that
the $^{30}$Si + $^{238}$U and $^{36}$S + $^{238}$U reactions
at low incident energies are governed by significantly different reaction processes, that are
reflected in the peaks of MDFF.
For the former reaction, the peak at $A \sim 178$ (the asymmetric peak)
originates from QF, while the peak at $A \sim 135$ (the symmetric peak) originates both from DQF and FF.
On the other hand, the mass-asymmetric peak for the latter reaction is located at $A \sim 200$,
originating predominantly from QF.

\section{Summary}

We developed a new dynamical model to describe heavy-ion induced fission, in which the effects
of static nuclear deformation of a target nucleus are taken into account by considering all
the orientation angles of
the symmetry axis of the target nucleus.
The orientation effects are included both in the barrier penetration process
and in the evolution of the nuclear shape.
The former process is described with the coupled-channels model.
After the nuclear contact point, we switch to the dynamical calculation
starting at the touching point assuming a nose-to-nose configuration.
The angle dependent touching distance $z_{touch}$ is introduced in order to effectively simulate the orientation effects
in the evolution phase of the nuclear shape.
With this model, the calculation could be extended to energies below the Coulomb barrier
for the first time.

In spite of the simplified assumptions, the calculation reproduced the measured
MDFF for the reactions of $^{36,34}$S+$^{238}$U and $^{30}$Si+$^{238}$U,
where a large variation of the distribution with respect to the incident beam energy are experimentally
observed.
By analyzing the Langevin trajectories, we could distinguish three different fission processes,
QF, DQF and FF.
The mass-asymmetric fission in $^{36}$S+$^{238}$U at $A_{L}/A_{H} = $ 74/200
observed at low incident energies are from QF.
The mass-asymmetric fission with $A_{L}/A_{H} = $ 90/178 in $^{30}$Si+$^{238}$U
observed at sub-barrier energy is also from QF, whereas the peak at the symmetric fission
of $A$=135 indicates that the fission occurs with some time delay and originates from FF and DQF.
The calculation suggests the formation of a mono-nucleus with a small $z$ value but with a
large collective fluctuation on the deformation axis of $\delta$.
A relatively long-life mono-nucleus is formed by the potential pocket
appearing inside the ridge line.
During the process of surmounting the ridge line, the mass asymmetry ($\alpha$)
of the system moves to the symmetric region.
This can be an account for the measured mass asymmetry in QF in $^{30}$Si+$^{238}$U.
The trajectories for $^{36}$S+$^{238}$U can not cross the ridge line and
are directed to QF without forming the mono-nucleus, thus the
mass-asymmetry is close to that of the entrance channel.

In this model, one can determine the fusion probability by selecting trajectories
which enter the fusion box.
The reproduction of the experimental MDFF in this model can be the ground to
support the calculated fusion probability.
Furthermore, the generalized formula proposed in this model has a potential to simulate
any kind of heavy-ion induced reactions in the approaching phase, such as a nucleon-transfer reaction,
and to predict cross sections for the production of new nuclei.



\section*{Acknowledgments}

The authors are grateful to Prof. V.I.~Zagrevaev, Prof. W.~Greiner,
Dr.~A.V.~Karpov, Dr.~F.A.~Ivanyuk, and Prof. F.~Hanappe
for their helpful suggestions and valuable discussions.
Special thanks are due to Dr. A.K.~Nasirov and Dr. A.S.~Denikin
for their helpful comments, and to
the JAEA experimental group for providing us with
the experimental data and for their helpful discussion.
The diabatic and adiabatic potentials were
calculated using the NRV code \cite{zagr07}.
The numerical calculations were carried out on SX8 at YITP at Kyoto University.
This work was supported by the Japanese
Ministry of Education, Culture, Sports, Science and Technology
by Grant-in-Aid for Scientific Research under
the program number  (C) 22540262.

\vspace{10cm}
\newpage
\clearpage

\vspace{10cm}
\newpage
\clearpage

\end{document}